\documentclass[preprint,aps,prl,floatfix,showpacs]{revtex4}
\usepackage{graphicx}

\begin{document}
\title{
A phase diagram for band inversion of topological materials as a function of interactions 
between two involved bands
}

\author{Jie-Xiang Yu}
\author{J. G. Che}\altaffiliation{Corresponding author.
E-mail: jgche@fudan.edu.cn.}
\affiliation{Surface Physics Laboratory (National Key Laboratory),
Key Laboratory of Computational Physical Sciences (MOE)
and Department of Physics, Fudan University,
Shanghai 200433, People's Republic of China}


\begin{abstract}
Based on first principles calculations, we predicate that Bi on a graphene derivate, 
$g$-C$_{14}$N$_3$, which involves a $3\times 3$ unit cell of graphene with four C atoms 
substituted by three N atoms, is a topological insulator with a gap of 50~meV. With the help of 
maximally localized Wannier functions, we find that the band inversion gap can be determined by 
examining a pair of interaction parameters between the two involved bands. Accordingly, a phase 
diagram for band inversion of topological materials as a function of the 
interactions is obtained. The conclusion also holds for Sb, Ir and Rh on $g$-C$_{14}$N$_3$. 
These materials are topological nontrivial either insulator or semimetal, indicating that 
$g$-C$_{14}$N$_3$ is a good platform for conceiving topological materials.
\end{abstract}

\maketitle
%motivation
A topological insulator (TI) exhibits a novel state that possesses simultaneously insulating bulk 
and conducting surface (edge) in one material. This is impossible to achieve in conventional 
materials.\cite{Has10,Qi11} The exotic physical properties of TI has potential applications in 
quantum devices and spintronics.\cite{Teo10,Lut10,Fu08,Yu10,Zha13} There are three keys to 
realizing a TI: time-reversal symmetry (TRS), spin-orbit coupling (SOC) and band inversion 
(BI).\cite{Has10,Qi11} TRS and SOC are dependent on atomic configuration and the atom 
properties themselves. Therefore, the primary challenge in the field falls centrally in BI, which can 
be manipulated by applying stress~\cite{Ber06,Kon07,Chu13,Pal14} and controlling the alloy 
component.\cite{Hsi08,Zha09} Recent research has suggested that a distortion in metal 
dichalcogenides could cause an intrinsic BI between chalcogenide-$p$ and metal-$d$ bands to 
ultimately form TI.\cite{Qia14} 
Applying stress, controlling the alloy component and distorting the lattice can affect 
the relevant interactions. However, little research to date has examined how BI depends on these interactions, 
and this is of direct importance in the application of TI.

%main finding
In the present work, 
we study the BI dependence on the interactions of the 
involved bands for Bi on $g$-C$_{14}$N$_3$ and present a phase diagram for the BI-gap as a 
function of interactions between two involved bands. The phase diagram works also for Sb, Ir and 
Rh on $g$-C$_{14}$N$_3$. 
%methods
The results were obtained by performing first principles calculations in the framework of density 
function theory with the projector augmented plane-wave (PAW) pseudopotential 
method,\cite{PAW} as implemented in VASP package.\cite{VASP} The exchange-correlation 
potentials were described by the local density approximation (LDA),\cite{LDA} which well 
described interactions in the graphene-based structures.\cite{Wang09} The energy cutoff for the 
plane-wave basis was 600 eV throughout all calculations. The k-points in the 2D Brillouin zone 
(BZ) of the $3\times 3$ size unit cell of graphene were sampled on a g-centered $7\times 7$ mesh. 
All atoms could be relaxed until the Hellman-Feynman forces on the atoms were less than 
0.001~eV/\AA. The SOC interaction was included. Spin-polarization was performed in the 
calculations. However, no spin-polarization effects were identified for the systems concerned. 
Maximally localized Wannier functions (MLWF) methods were implemented in the Wannier90 
package for tight-binding parameters.\cite{MLWF}

\begin{figure}[bt]
%\centerline{\includegraphics[scale=0.70,angle=0]{config.eps}
%\centerline{\includegraphics[scale=1.20,angle=0]{config.eps}
\centerline{\includegraphics{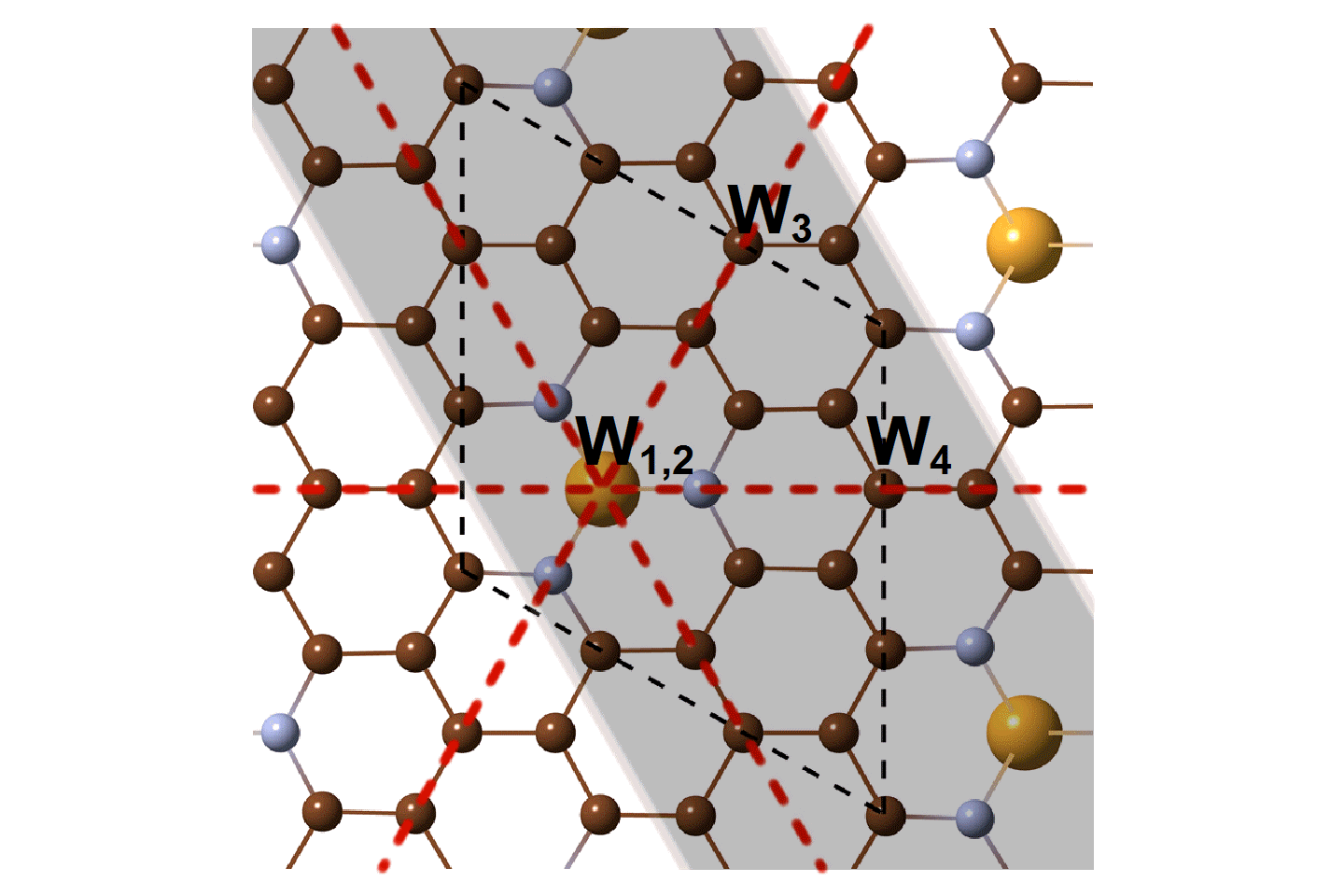}}
\caption{(Color online)
Topview of the atomic configuration for Bi(Sb, Rh, Ir) on $g$-C$_{14}$N$_3$. 
Brown, blue and yellow balls represent C, N and Bi(Sb, Rh, Ir), respectively. Thin dashed lines 
denote the boundary of a $3\times 3$ unit cell of graphene with three mirror-symmetric planes, 
which are identified by three thick red dashed lines. The shadow region indicates a 1D-ribbon 
with an armchair edge. $W_1$, $W_2$, $W_3$ and $W_4$ indicate Wannier function centers 
optimized by the MLWF method. 
}
\label{config}
\end{figure}

%atomic configuration of Bi/$g$-C$_{14}$N$_3$
Firstly, we focused on Bi on $g$-C$_{14}$N$_3$. The most stable structure of 
Bi/$g$-C$_{14}$N$_3$ is quite similar to that of Au/$g$-C$_{14}$N$_3$~\cite{Yu15} and the 
topview of its atomic configuration is shown in Fig.~\ref{config}. Bi was lifted by 1.66~\AA\ 
away from the graphene layer and the shift of three N atoms upwards to the graphene layer is 
smaller than 0.01~\AA. Each N is bonded to two neighbor C atoms with the length of C-N being 
1.37~\AA. At 123$^o$, the C-N-C angle is slightly larger than that of the C atoms in graphene 
(120$^o$). Bi on $g$-C$_{14}$N$_3$ obtains about an energy of 3.0~eV, meaning that 
%this atomic configuration of 
Bi on $g$-C$_{14}$N$_3$ is quite stable.

% features of $g$-C$_{14}$N$_3$ 
Before discussing the electronic properties of Bi/$g$-C$_{14}$N$_3$ in more detail, it is worth 
briefly reviewing the features of the $g$-C$_{14}$N$_3$ band structure that were described in 
depth in our previous work.\cite{Yu15}. The $g$-C$_{14}$N$_3$ keeps the main electronic 
feature of graphene, the Dirac-point with a small gap of 0.1~eV between the $\pi$ and 
$\pi$* band. However, since four C atoms are replaced by three N atoms, the Fermi level lies 
0.2~eV below the top of the $\pi$ band, resulting in a $+1e$ hole. The intercation bands 
between three N lone pairs are fully occupied and the Fermi level lies at the band top. 
The $\pi$' band, consisting of about 25\% N(pz) and 75\% C(pz), 
degenerates with the $\pi$ band at the $\Gamma$ point. 
Its antibonding band, $\pi$'*, is about 2.5~eV above the Fermi level.
Near the M point, there exists a crossing of the $\pi$* and $\pi$'*, 
which plays an important role in forming a TI when Bi 
deposited on $g$-C$_{14}$N$_3$. In Au/$g$-C$_{14}$N$_3$, one valence electron of Au fills 
the $+1e$ hole, resulting in the Fermi level returning to the Dirac point. Accordingly, an ion 
with two valence electrons more than that of Au adsorption on $g$-C$_{14}$N$_3$ could shift the 
Fermi level to the crossing of $\pi$* and $\pi$'*,
since the $\pi$* band is the lowest conducting 
band of $g$-C$_{14}$N$_3$. Bi in an $s^2p^3$ configuration is this type of ion due to its two 
$s$-electrons in a deep level of about $-7\sim-8$~eV.

\begin{figure}[bt]
%\centerline{\includegraphics[scale=0.70,angle=0]{Band.eps}}
%\centerline{\includegraphics[scale=1.20,angle=0]{Band.eps}}
\centerline{\includegraphics{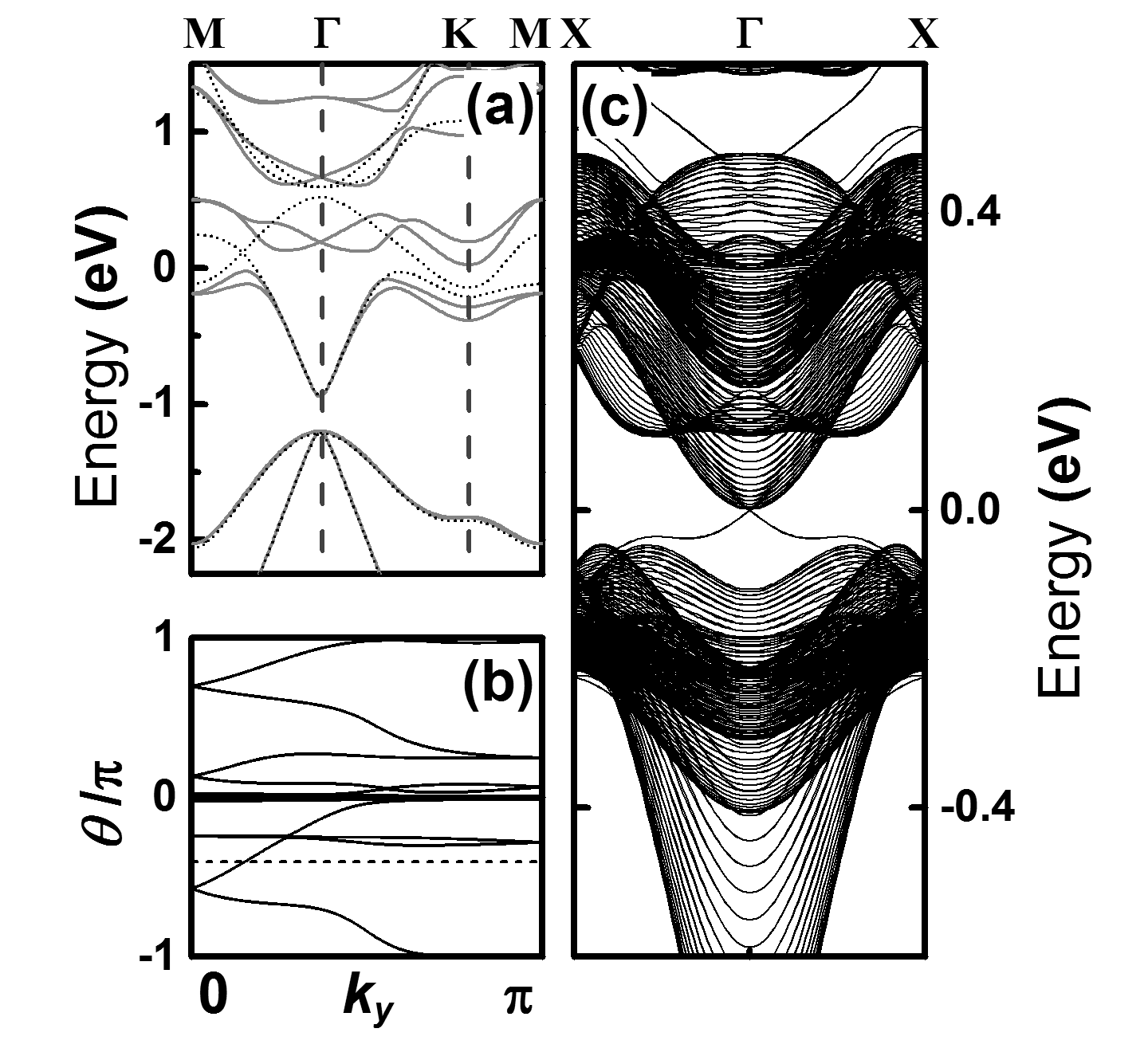}}
\caption{
(a) Band structure of Bi on $g$-C$_{14}$N$_3$ with (solid lines) and without (dotted 
lines) spin-orbit coupling, (b) Evolution of Wannier function centers
for Bi on $g$-C$_{14}$N$_3$, and (c) 1D 
band structure of Bi/$g$-C$_{14}$N$_3$ stack-ribbon. 
}
\label{Band}
\end{figure}

% features of Bi/$g$-C$_{14}$N$_3$
The band structures of Bi/$g$-C$_{14}$N$_3$, along the M, $\Gamma$, K and M point with 
(solid lines) and without SOC (dashed lines), were calculated and are shown in Fig.~\ref{Band} 
(a). The band structure without SOC (dashed lines) reveals that two bands cross over the Fermi 
level near the M-point as well as near K-points, indicating that the system without SOC is metallic. 
Furthermore, the two bands cross between M-G, which is referred to as $\Lambda$. 
These two bands can be traced to 
the empty $\pi$* and $\pi$'* bands of $g$-C$_{14}$N$_3$, occupied by $p$-electrons of Bi. The 
$\Lambda$-point in Bi/$g$-C$_{14}$N$_3$ can also be traced to the crossing of the empty $\pi$* and 
$\pi$'* bands in $g$-C$_{14}$N$_3$. However, in Bi/$g$-C$_{14}$N$_3$, the $\Lambda$-point is 
1.0~eV above the Dirac point, while in $g$-C$_{14}$N$_3$ the crossing is 2.5~eV above the 
Dirac point. 
Taking SOC into 
account, the band structures (solid lines) exhibit a strong SOC splitting, a gap of 145~meV opened 
surrounding the $\Lambda$-point. Furthermore, the mixed $\pi$'* 
band, which is occupied around the K-point without SOC, shifts up 160~meV and becomes 
unoccupied, while the mixed $\pi$* band shifts down 66~meV. By shifting in opposite direction, 
the two bands open a gap of 50~meV in the axis $\Gamma$-K-M, which is overlapped by the gap 
surrounding the $\Lambda$-point and is thus an optical gap. The Fermi level lies in the gap. 

In order to check whether the gap is topological nontrivial or not, we calculated the evolution of 
WFC~\cite{EoWFC} and electronic structures for a stacked-ribbon with 50 1D ribbons of 
Bi/$g$-C$_{14}$N$_3$. The results are shown in Fig.~\ref{Band} (b) and (c), respectively. 
Both the odd winding number of WFC and gapless edge states indicate that the gap surrounding 
the $\Lambda$-point is topological nontrivial.

%building H matrix
%concentrated on the involved bands
Along the axis $\Gamma$-K-M 
of BZ of Bi/$g$-C$_{14}$N$_3$, the mixed $\pi$'* band is higher than the mixed $\pi$* band, 
regardless of whether SOC was included (solid lines) or not (dotted lines) in calculations. This 
indicates that no bands are inverted along the $\Gamma$-K-M. Considering the importance of BI, 
which occurs at the M-point of BZ and 0.1~eV above the Fermi level for Bi/$g$-C$_{14}$N$_3$ 
without SOC, as indicated by dotted lines in Fig.~\ref{Band} (a), we then turned our attention to 
identifying which factors determine the BI. 

%Bi as an ion of Bi+3
The SOC effects of Bi/$g$-C$_{14}$N$_3$, which can be traced to the contribution of 
$p$-electrons of Bi$^{+3}$, were obtained to compare the band structures presented by the dotted 
(without SOC) and solid (with SOC) lines. As the calculated electronic structures indicate, the 
SOC effect mainly exists in the bands above -1.0~eV, as shown in Fig.~\ref{Band} (a). In 
contrast, the difference in the solid and dotted lines below -1~eV indicates less SOC splitting on 
these bands. This is because three of five Bi valence electrons fill the bands below -1~eV: one 
electron in the $+1e$ hole of the $g$-C$_{14}$N$_3$~\cite{Yu15} centered at about -1~eV, and 
the other two in the $s$-level within $-7\sim-8$~eV. The bands above -1.0~eV are fully separated by a 
gap of 0.3~eV from the bands below, as shown in Fig.~\ref{Band} (a). The remaining two 
electrons of ion Bi$^{+3}$ in a doublet $p_x+p_y$ configuration exhibit atomic-like behaviors 
within a rhombohedral crystal field and fill in the band above -1~eV. Thus, the bands above -1~eV 
could be seen as those involved in ion Bi$^{+3}$ with strong SOC. Accordingly, we focused only 
on the bands above -1~eV and will analyze these in more detail below.

\begin{figure}[bt]
%\centerline{\includegraphics[scale=0.60,angle=0]{BI.eps}}
%\centerline{\includegraphics[scale=1.10,angle=0]{BI.eps}}
\centerline{\includegraphics{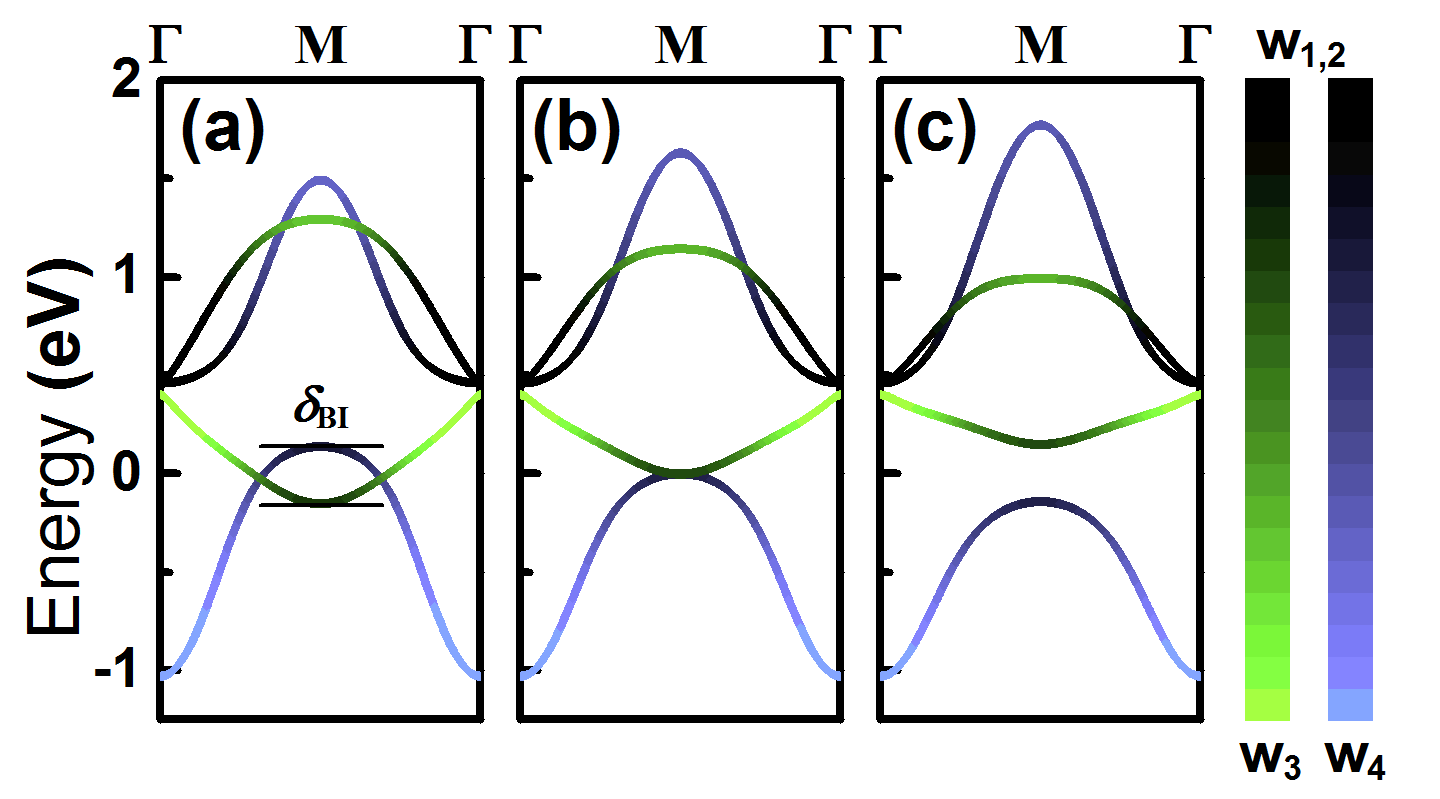}}
\caption{(Color online)
Band structures for BI relative to $t_1/t_2$ parameters. The colors indicate the 
corresponding contributions of WF, given in the color scale. 
The corresponding parameters in Eq.~\ref{delta} 
are taken as $a = -1.04$, $b = -0.91$, $c = -0.90$, $\alpha=0.04$, $\beta=-0.06$, 
$\gamma=-0.54$ (in eV). 
}
\label{BI}
\end{figure}

%MLWF for TB parameters
During the optimizing process that was designed to transform the involved bands of 
Bi/$g$-C$_{14}$N$_3$ above -1~eV from our first principles calculations to the MLWF, we 
chose an inner energy-window of $-1.00\sim 0.55$~eV including the involved bands across the Fermi 
level. Thus, the MLWF was performed for 12 bands above -1~eV as an outer energy-window. 
During the process, four Wannier functions were chosen, two ($W_1$ and $W_2$) for the Bi's 
$p_x/p_y$ orbital due to its three-fold coordination, and the other two ($W_3$ and $W_4$) for the 
$\pi$* orbital and the $\pi$'* orbitals respectively. The optimized centers of four WF, as shown in 
Fig. 1 and indicated by the labels $W_1$, $W_2$, $W_3$ and $W_4$, are independent on their 
initial positions.

%building H
We then used the parameters obtained from the optimizing process to construct the Hamiltonian 
along axis $\Gamma$-M. In this construction, we ignored interactions with distances larger than third 
nearest-neighbors, corresponding to the interactions up to the involved WF centers over the unit 
cell. Despite the truncation of the interactions, the main features of these bands remained. After a 
rotation transformation 
%of defined by a matrix of 
%\begin{eqnarray*}
%\left( \begin{array}{cc}
%\cos \theta & -\sin \theta\\
%\sin \theta & cos \theta \\
%\end{array} \right) 
%\end{eqnarray*}
%
%\noindent 
with $\theta = \pi/6$, the Hamiltonian matrix

\begin{eqnarray*}
\label{eqH}
\left( 
\begin{array}{cccc}
{a+\alpha(1-\cos k)} &0    &{2t_1(1-e^{-ik})}        &0    \\
0      &a+\alpha(1-\cos k) &0          &{-2t_2(1-e^{ik})}  \\
2t_1(1-e^{ik})       &0    &b+\beta(2\cos k+\cos 2k) &0    \\
0      &-2t_2(1-e^{-ik})   &0          &c+\gamma(1+2\cos k)\\
\end{array} 
\right),
\end{eqnarray*}

\noindent 
could be obtained. Here $a$, $b$, $c$, $\alpha$, $\beta$, $\Gamma$, $t_1$, and $t_2$ were the 
interaction parameters obtained from the MLWF optimizing process. This matrix had a form of 
two $2\times 2$ diagonal submatrixes. Thus, with $A=a+2\alpha$, $B=b-\beta$, $C=c-\gamma$, one 
could easily obtain the BI gap.\cite{Qia14} as

\begin{equation}
\label{eqBI}
\delta_{\rm BI}=C-B-\sqrt{(A-C)^2+64 t_2^2}+\sqrt{(A-B)^2+64 t_1^2}.
\label{delta}
\end{equation}

\noindent 
Since $a$, $b$, $c$, $\alpha$, $\beta$, $\Gamma$ in 
Hamiltonian are on-site and self-interaction parameters, the BI were determined by the $t_1$ and 
$t_2$ parameters, which describe the interactions for $W_1$-$W_3$ and $W_2$-$W_4$ 
respectively. The $t_1/t_2$ optimized by the MLWF method for Bi/$g$-C$_{14}$N$_3$ reads 
0.180/0.150, corresponding to a BI-gap of 0.29~eV, as shown in Fig.~\ref{BI} (a). $t_1/t_2$ can 
be used to tune the BI gap: if $t_1/t_2$ is taken as 0.142/0.188 and 0.104/0.226, the BI-gap is zero 
and -0.29~eV, respectively, as shown in Fig.~\ref{BI} (b) and (c).

\begin{figure}[bt]
%\centerline{\includegraphics[scale=0.60,angle=0]{SbIrRh.eps}}
%\centerline{\includegraphics[scale=1.10,angle=0]{SbIrRh.eps}}
\centerline{\includegraphics{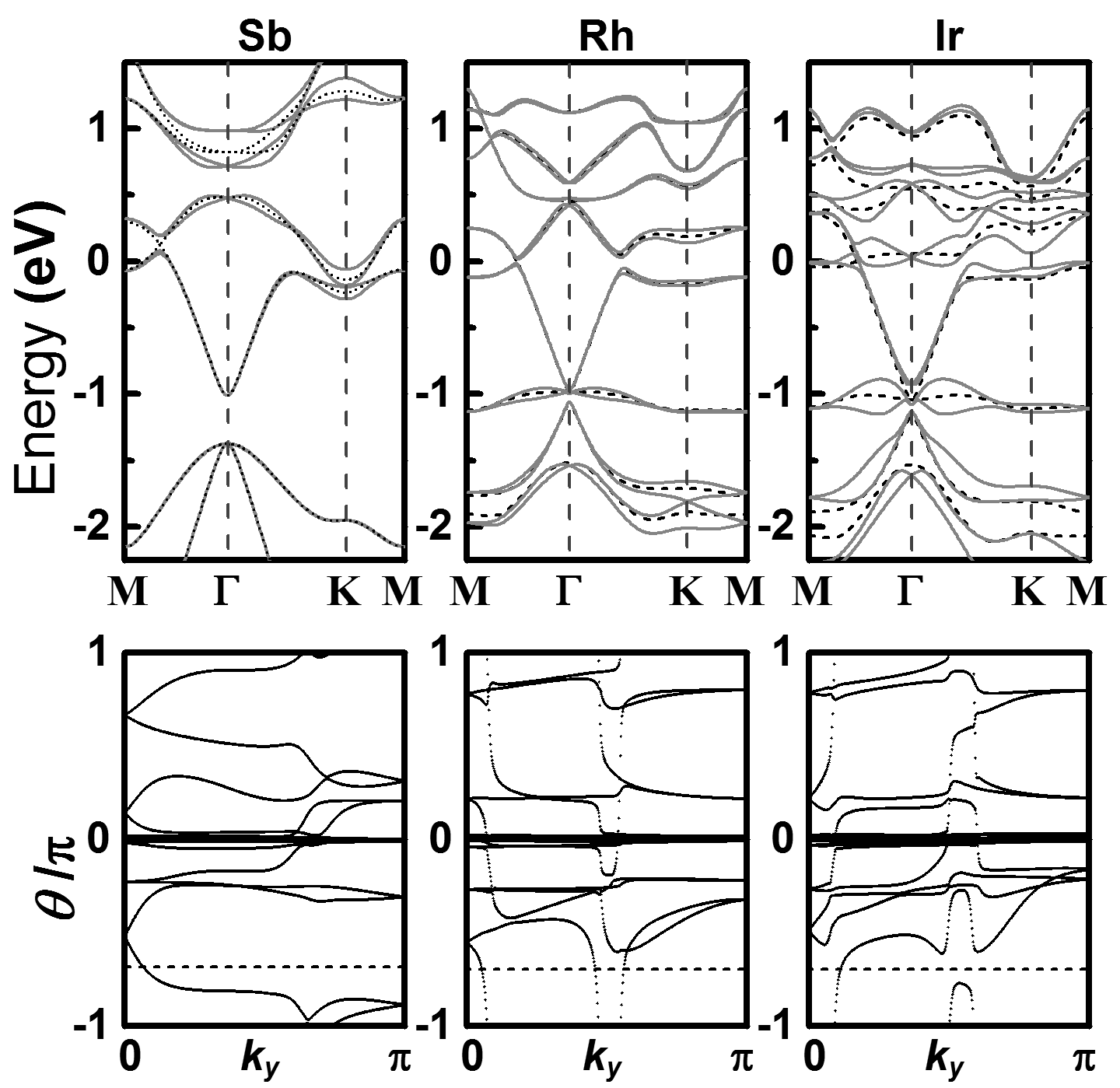}}
\caption{
Electronic structures of Sb, Rh and Ir on $g$-C$_{14}$N$_3$ labled by Sb, Rh and Ir 
respectively. The up and down panels correspond to the band structures and evolution of WFC.
}
\label{SbIrRh}
\end{figure}

%results of Sb, Rh and Ir
The electronic structures of Sb, Rh and Ir on $g$-C$_{14}$N$_3$ were obtained by the same 
processes and are indicated in Fig.~\ref{SbIrRh} by Sb, Rh and Ir labels respectively. In the figure, 
the up and down panels correspond to the band structure and the evolution of WFC respectively.
The evolutions of WFC for all three cases exhibit an odd winding 
number, indicating a topological nontrivial gap along the M-$\Gamma$ axis. However, due to the 
fact that they exhibit a different band structure along the $\Gamma$-K axis, the case of Sb and Ir 
on $g$-C$_{14}$N$_3$ are topological semimetal, whereas the Rh case is a topological 
nontrivial insulator with a gap of 30~meV.

\begin{figure}[bt]
%\centerline{\includegraphics[scale=0.70,angle=0]{phase.eps}}
%\centerline{\includegraphics[scale=1.20,angle=0]{phase.eps}}
\centerline{\includegraphics{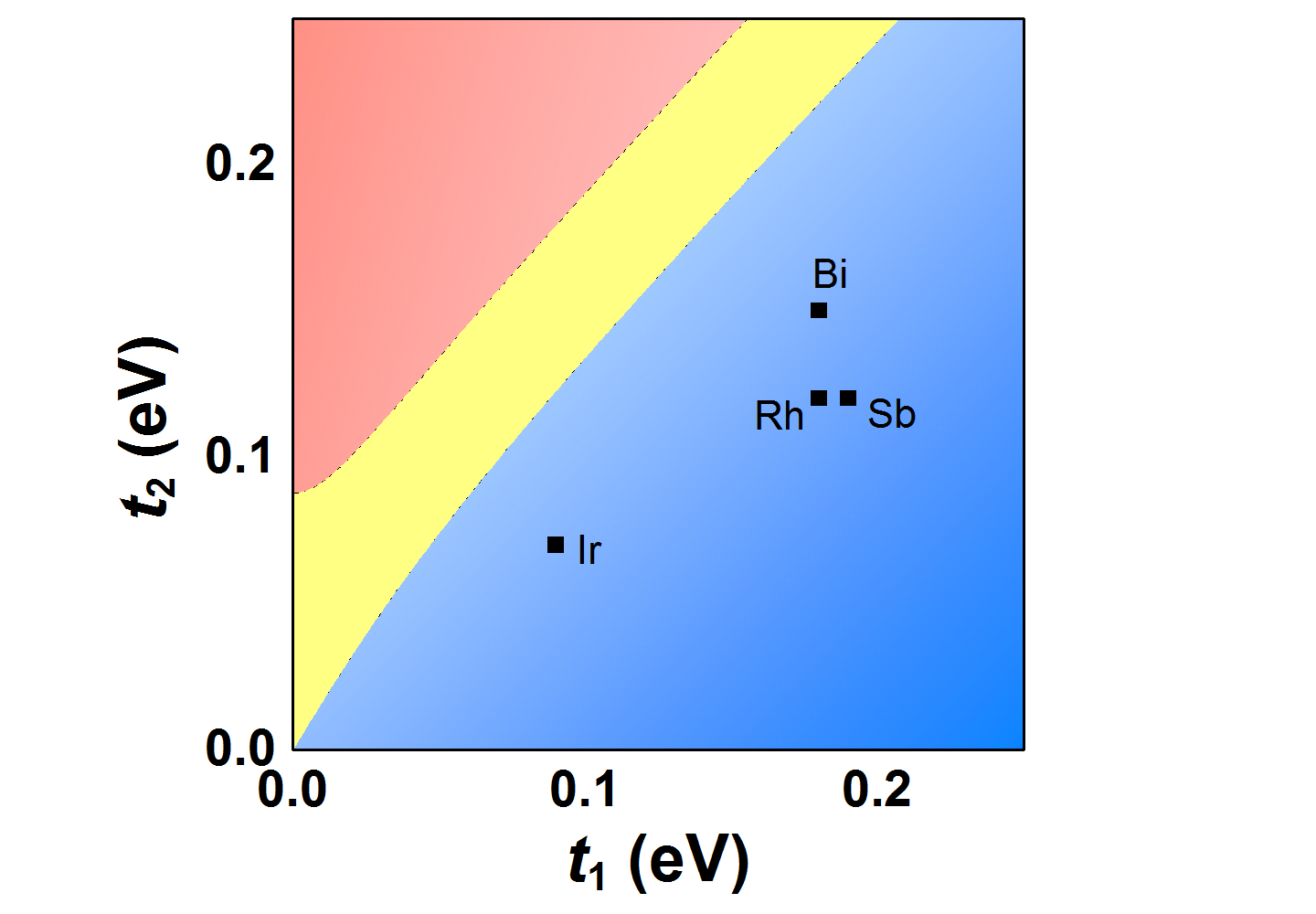}}
\caption{(Color online)
Phase diagram of the band inversion gap as a function of interaction parameters $t_1/t_2$. 
The blue and red regions correspond to topological nontrivial and trivial phase respectively. The 
Bi, Sb, Rh, and Ir labels indicate the optimized parameter pairs of $t_1/t_2$ for the case of Bi, Sb, 
Rh and Ir on $g$-C$_{14}$N$_3$ respectively. The yellow region corresponds to either positive 
or negative BI-gap, case by case.
}
\label{phase}
\end{figure}

Through applying similar processes to the case of Bi on $g$-C$_{14}$N$_3$, we also obtained 
the optimized parameters in Eq.~\ref{delta} for cases of Sb, Rh and Ir on $g$-C$_{14}$N$_3$; 
that is, the BI-gap as a function of the interaction between the involved bands for Sb, Rh and Ir on 
$g$-C$_{14}$N$_3$. Eq.~\ref{delta} indicates that positive, zero and negative values for the 
BI-gap could be tuned by pairs of the interaction parameters $t_1/t_2$. Accordingly, a phase 
diagram of the BI as a function of the interaction parameters between two involved bands could be 
obtained. This is shown in Fig.~\ref{phase}, in which the red region indicates a topological trivial 
phase, while the blue region indicates a topological nontrivial phase. The optimized interaction 
parameters $t_1/t_2$ for Bi, Sb, Rh and Ir on $g$-C$_{14}$N$_3$ all lie in the blue region (with 
positive BI-gap), labeled by Bi, Sb, Rh and Ir respectively, revealing that all four cases have a 
topological nontrivial gap. This indicates that $g$-C$_{14}$N$_3$ is a good platform for 
conceiving topological nontrivial materials. 

%conclusions
In conclusion, we demonstrated that Bi/$g$-C$_{14}$N$_3$ is a topological insulator with a 
topological nontrivial gap of 145~meV and an optical gap of 50~meV. The study also confirmed 
that depositing Sb (Ir and Rh) on $g$-C$_{14}$N$_3$ could also form topological nontrivial 
materials (either insulator or semimetal). These results show that $g$-C$_{14}$N$_3$ is a good 
platform to conceive a TI: an ion with three valence electrons with strong SOC on the platform 
could lead to the Fermi level shifting to the $\Lambda$-point and opening a topological nontrivial gap. 
Based on MLWF our analysis indicated that the band inversion could be determined by 
interactions between two involved bands. Accordingly, a phase diagram of the band inversion gap 
as a function of the interaction parameters is presented.

%acknowledgments
This work was supported by NFSC (No.61274097) and NBRPC (No. 2011CB921803 and 2015CB921401).

\end{document}